\documentstyle[a4,12pt]{article}
\begin{document}
\voffset 0.6 cm
\hoffset 0.25in
\setcounter{page}{1}
LPTHE Orsay-93/33
06 Sept 1993.
\newcommand{\be}{\begin{equation}}
\newcommand{\ee}{\end{equation}}
\newcommand{\bea}{\begin{eqnarray}}
\newcommand{\eea}{\end{eqnarray}}
\newcommand{\nn}{\nonumber}
\newcommand{\muh}{\hat\mu}
\newcommand{\dlr}{\stackrel{\leftrightarrow}{D} _\mu}
\newcommand{\vnew}{$V^{\rm{NEW}}$}
\newcommand{\vecp}{$\vec p$}
\newcommand{\dof}{{\rm d.o.f.}}
\newcommand{\prd}{Phys.Rev. \underline}
\newcommand{\pl}{Phys.Lett. \underline}
\newcommand{\prl}{Phys.Rev.Lett. \underline}
\newcommand{\np}{Nucl.Phys. \underline}
\newcommand{\vvp}{v_B\cdot v_D}
\newcommand{\dl}{\stackrel{\leftarrow}{D}}
\newcommand{\dr}{\stackrel{\rightarrow}{D}}
\newcommand{\mev}{{\rm MeV}}
\newcommand{\gev}{{\rm GeV}}
\newcommand{\calp}{{\cal P}}
\newcommand{\ra}{\rightarrow}
\pagestyle{empty}
\vskip 1cm
\centerline{\Large{\bf{SEMILEPTONIC DECAYS AND CKM}}}\vskip 0.3 cm
\centerline{\Large{\bf{ MATRIX ELEMENTS IN A TCF}}}
\vskip 0.5 cm
{Talk delivered at the Marbella TCF workshop,1-6 june 1993}
\vskip 1.5cm
{\it{A.Le Yaouanc and O.P\`ene.}}
\vskip 0.0cm
{ LPTHE, Orsay, France\footnote {Laboratoire associ\'e au
 Centre National de la Recherche Scientifique.}.}
\begin{abstract}
\narrower
\noindent
We argue that a Tau-Charm factory (TCF) could improve our knowledge
 of CKM matrix elements since it is an incomparable tool to check the models
and methods applied to extract $V_{bu}$ from $B$ decay partial widths.
 We report on some recent proposals to improve
 on parton model. Turning to exclusive
decays, we compare the predictions from quark models, QCD sum rules,
effective Lagrangians and lattice QCD. Quark
models have anticipated on heavy quark symmetry.  Their difficutly
to account for the $q^2$ dependence might be partly cured by relativistic
corrections.  QCD  sum rules
and lattice seem to disagree on the $q^2$ dependence of axial form factors.
We discuss the extrapolation from $D$ to $B$. Present uncertainties
do not allow to predict reliably the  $B\ra\pi,\rho l \nu$
 matrix elements.  We argue that QCD sum rules are
in a good position to study $q^2$ dependence down to $q^2=0$. Lattice
QCD is able to check the
heavy quark scaling laws for the heavy to light semileptonic decays.
It seems
to confirm the increase of $A_2/A_1$ from $D$ to $B$ meson predicted by the
scaling laws.
Finally  semileptonic decays in a TCF would give incomparable informations
on $K\pi$ phase shifts, and on $K^{**}$ resonances.
\end{abstract}
\pagestyle{plain}
\section{INTRODUCTION}
\label{sec:intro}
During our working group on CKM matrix elements and semileptonic decays,
we heard five interesting talks. Four of them were
dealing mainly with heavy flavours semileptonic decays: by P. Roudeau on
experimental results, by P. Colangelo on QCD sum rules, by V. Lubicz on lattice
QCD and by N. Di Bartolomeo on an effective Lagrangian approach. The fifth
talk, by M. Shifman, was more generally advocating a new operator expansion
method for heavy flavours. We will try to summarise these enlightening
contributions, and add some comments on quark model studies of heavy flavours
semileptonic decays.
Semileptonic decays of heavy mesons have attracted considerable
interest in the past years as they play a crucial role in the determination
of the Cabibbo-Kobayashi-Maskawa mixing matrix. Now the question is: what could
a Tau-Charm factory (TCF) learn us about CKM angles ? At first sight the answer
seems rather negative. A TCF would mainly give a direct access to $V_{cs}$ and
$V_{cd}$. It is true that these angles are not directly known to a high
degree of
accuracy, but unitarity constrains them very strongly from $V_{us}$ and
$V_{ud}$. Nobody would bet a cent on the chances of something new to happen
here.
It is well known that the CKM angles we are mostly interested in, are those of
the
third generation. And to know them we need a $b$-factory. What is then the use
of
a TCF in that respect ? The fact is that a $b$-factory gives you an
experimental
number, say a partial width, $\Gamma = \vert V_{ub}\vert^2  X$, where $X$ is,
 up to kinematical factors, the squared matrix element of some operator between
hadrons. $X$ is not given by any symmetry principle (except in one point of
phase space), $X$ is in general a very difficult quantity to compute
theoretically. Any model, any theoretical method that makes predictions about
the
$B$ decay matrix elements has something to say also about charm decay and can
be checked there since we know the relevant CKM angles. This point is crucial,
{\it a good understanding of charm physics is an unescapable necessity to make
reliable predictions concerning beauty}. The belief that heavy Quark Effective
Theory (HQET) could save us this effort is far too naive. Heavy Quark
Symmetry is a very useful tool, but it does not provide us with the needed
quantitaties (we will see some examples in the following). It is also
fair to say that the extrapolation from charm to beauty is by no means trivial
and needs a lot of theoretical work.
One major theoretical challenge to physics today is to
improve our understanding of the non-perturbative aspects of QCD. M. Shifman
expressed the view that a TCF would be a new and powerful microscope into
strong
interactions. Indeed the round table of this workshop has stressed that a TCF
could increase dramatically the accuracy of experimental data related to
non-perturbative QCD. Mutatis mutandis, this could be compared to the role of
LEP as related to the perturbative aspects of the standard model. Many
illustrations
of what a TCF could bring in this respect have been produced during this
workshop.
 The precise measurement of hadronic matrix element of
$\Delta C=1$ currents, that will be achieved from semileptonic decays, is
another example of what a TCF can learn us about non-perturbative QCD.
 Finally, although this is not too popular at present, we should not forget
that the analysis
of non charmed final states of semileptonic decays would be a rich source of
informations.
\subsection{Vector meson dominance and scaling laws.}
\label{sub:nk}
{}From Lorentz invariance, it is possible to express in all generality the
current matrix elements as:
\be < K \vert J_\mu \vert D > = \Bigl( p_D + p_K -
\frac{ M_D^2 - M_K^2}{q^2} q \Bigr)_\mu f^+_K(q^2) +
\frac{ M_D^2-M_K^2}{q^2} q_\mu f^0_K(q^2) \label{dk}\ee
\bea
 < K^*_r \vert J_\mu \vert D > &=& e^\beta_r \Bigl[ \frac{2
V(q^2)}{M_D+M_{K^*}}
\epsilon_{\mu\gamma\delta\beta}p_D^\gamma p_{K^*}^\delta + i
( M_D +M_{K^*}) A_1(q^2) g_{\mu\beta} \nn \\ &-& i
\frac{A_2(q^2)}{M_D+M_{K^*}}P_\mu q_\beta + i \frac{A(q^2)}{q^2}2
M_{K^*}q_\mu P_\beta \Bigr] \label{dks} , \eea
where $q$ is the momentum
transfer, $q=p_D-p_K$ or $q=p_D-p_{K^*}$, $P=p_D+p_{K^*}$ and $e^\beta_r$ is
the polarization vector of the $K^*$.  $f^{+,0}_K$, $V$, $A_{1,2}$ and $A$ are
dimensionless form factors. For the axial current, some authors prefer
another set of form factors:
\bea f(q^2)=(M_D+M_K)A_1(q^2),&\,\,\nn\\
a_+(q^2)=-i\frac{A_2(q^2)}{M_D+M_K},\quad \quad &
a_-(q^2)=i\frac{A(q^2)}{q^2}.\label{ff2}\eea
Let us consider the kinematics of $P\ra M l \nu$, where $P$ is $D$ or $B$
and $M$ is a final light meson, $K$, $K^*$, $\pi$ or $\rho$.
 The physical region for semileptonic decay corresponds to
$0\le q^2\le q^2_{max}$ where $q^2_{max}=(M_P-M_M)^2$.
 $q^2_{max}$ corresponds to the no recoil point, i.e. the final meson as well
as the dilepton system are at rest in the initial restframe\footnote{All over
this report we will chose as reference frame
the rest frame of the initial heavy meson.} : $\vec q=0$.
This point is the closest to the nearest $t$-channel pole, located at
$q^2=M_t^2$
where $M_t$ is the mass of the lightest meson with the quantum numbers
exchanged in the $t$-channel. For example when the final meson is $K$ or $K^*$,
the exchanged
 meson is a $c\bar s$ meson, vector, axial or scalar meson according to the
cases.
 It is known that $M_t-M_P\ra \hbox{constant}$ when
$m_Q\ra\infty$\footnote{$m_Q$ is
the heavy quark mass.}. It follows
$M_t^2> M_P^2 > q^2_{max}$, but
$M_t-\sqrt{q^2_{max}}$ is not too large and stays constant when the
$M_P\ra\infty$. As a consequence, {\it the region near $q^2_{max}$ may
feel strongly the influence of the nearest pole}\footnote {This is usually
phrased,
 in the case of the form factor $f^+$, as the
vector meson dominance (VMD) hypothesis.}.
On the contrary, when $q^2$ decreases,
the distance to the nearest pole increases, and the relative influence of all
the
other singularities in the $t$-channel increases, {\it washing out the
dominance
of the nearest pole}.
Isgur and Wise\cite{sliw} have proposed a set of scaling laws relating
semileptonic amplitudes for a given final light meson ($K, K^*$ or $ \pi,
\rho$) when the mass of the initial meson goes to infinity:
 up to $O(1/M_P^2)$, up to logarithmic
corrections, one expects the following behaviour for the relevant form
factors\cite{sliw}:
\bea \frac{f^+}{M_P^{1/2}}&=& \gamma_+ \times \Bigl( 1+ \frac{\delta_+}{M_P}
\Bigr) \,\,\,\,\,\,\,\,\,\,\,\,\,
\frac{V}{M_P^{1/2}} = \gamma_V \times \Bigl( 1+ \frac{\delta_V}{M_P}
\Bigr) \nonumber \\ \frac{A_2}{M_P^{1/2}}&=& \gamma_2 \times \Bigl( 1+
\frac{\delta_2}{M_P} \Bigr)
\,\,\,\,\,\,\,\,\,\,\,\,\,
A_1 M_P^{1/2}= \gamma_1 \times \Bigl( 1+ \frac{\delta_1}{M_P}
\Bigr) \label{scale} \eea
where $M_P$ is the mass of the initial heavy meson.
The expansions given in eqs.(\ref{scale}) become valid in the limit of large
$m_Q$, at fixed momentum $-\vec q$
 of the light meson (in the frame where the
heavy meson is at rest) and when $\vert \vec q \vert \ll m_Q \sim M_P$. The
above conditions are always satisfied for $q_{max}$, when the initial and
final mesons are both at rest.
 For $q^2=0$, $\vert \vec q\vert=(M_P^2-M_M^2)/(2 M_P)\sim M_P/2$.
 Assuming, for instance, that we acquire a good knowledge
of $D\ra \rho$, the scaling laws in eq. (\ref{scale}) will lead to a prediction
of
$B\ra \rho$ in the region $\vert \vec q\vert < M_D/2$.  This is only a small
region close to $q^2_{max}$ in the physical phase space for
$B\ra \rho$, very far from $q^2=0$, the
latter being the dominant contribution to phase space. The conclusion is that
{\it the
scaling laws are not enough} to allow an extrapolation from $D\ra\rho$ to $B\ra
\rho$. A further extrapolation in $q^2$ from the small recoil region down to
$q^2=0$ is necessary. And this cannot been done reliably from the simplest use
of nearest pole dominance, as we have argued in the preceeding paragraph. {\it
This extrapolation is a formidable challenge to theory in the coming years}.

\section{INCLUSIVE SEMILEPTONIC DECAYS}
\label{sec:incl}
At present the only positive experimental source of information concerning
$V_{ub}$ comes from inclusive semileptonic $B$ meson decay, since ARGUS and
CLEO do not agree on the existence of a positive signal in exclusive decays.
These inclusive decays are compared, either to the parton model, or to the sum
of exclusive decays in some exclusive model. We will discuss the exclusive
models later on. The parton model, relying on duality, directly predicts the
inclusive decays from an amplitude computed with free final quarks, including
perturbative QCD corrections. As a first approximation the parton model is
very good. However an improvement of its accuracy is not easy since its
perdictions depend very strongly on the $b$ quark mass, on the $B$ meson wave
function and because the corrections to duality are not known. One needs a
systematic approach able to improve from the parton model.
Such a systematic non perturbative approach has been proposed by Bigi, Shifman,
Uraltsev and
Vainshtein\cite{bsuv}. Their idea is a generalisation of operator expansion, it
amounts to expand the heavy quark weak interaction {\it in the background of
the light quark and gluon fields} into a sery of local operators. It is an
expansion
in $1/m_Q$ (inverse of the heavy quark mass), the non trivial contributions
starting at $(1/m_Q)^2$. This is a very appealing method, but up to now only
semi-quantitative. Tests are needed, and the charm sector seems the best place
since the $1/m_c^2$ corrections are large (maybe even too large in some cases).
However an unlucky feature is that, due to singular contributions, the operator
expansion fails near the end point of the electron spectrum. This happens to be
the point where the suppression
of the charm background allows an experimental observation of $b \rightarrow u$
semileptonic decay. These singularities also prevent to predict the energy
spectrum in the case of charm decay, only integrated quantities being
predictable. We are thus lead to turn toward exclusive theoretical
calculational methods in order to estimate what degree of accuracy may be
expected from them.
\section{QUARK MODELS}
\label{sec:quark}
It must first be emphasized that in contrast to lattice QCD (and partly to QCD
sum rules), quark models (QM) most often do not provide constraining
predictions. They are not definite approximations of QCD and their predictions
depend on various parameters (quark masses, potentials) which have no direct
physical meaning and make sense only within these specific models. These
parameters can be fixed independently through spectroscopy, but only roughly.
Exceptions to the lack of constraining predictions are for example the heavy
quark symmetry relations, but they have been shown to derive from much more
general principles.
Being interested in semileptonic decays, the main interest of quark models lies
in the {\it physical insight} they provide through the dynamical concepts of
composite systems. However the position and reliability of quark models is
 strongly dependent on the channel and decreases when going from $B\rightarrow
D, D^*$
transitions,
to  $D\ra K, K^*$ or $D\ra \pi,\rho$ decays, not to speak of $B\ra \pi,\rho$
which presents
very large momentum transfers.
\subsection{Quark models and Isgur-Wise relations.}
\label{sub:iw}
Quark models, even in a rather naive stage, give straightforwardly the
probable pattern of magnitude of the various form factors, which for the ground
state transitions $0^-\ra 0^-, 1^-$ already amount to six a priori independent
ones. This simple pattern has been first put forward by Altomari and
Wolfenstein\cite{aw} who have discovered the simplificatory virtues of
$\alpha$)
having a light spectator with heavy flavor transition, $\beta$) working near
$q^2_{max}$ (the no recoil point). Most predictions derive from the
non-relativistic quark model (NRQM). An exception is $A_2$ ($a_+$), crucial
for the polarization in $0^-\ra 1^-$ decays, which can be derived consistently
from a subtle treatment including Wigner rotations\cite{gilopr}. The general
predicted patern at $q^2_{max}$ is in {\it qualitative} agreement with what is
known experimentally \footnote{Unhappily, due to phase space, experimental
measurements are rather taken near $q^2=0$}. The agreement
seems now to include also $A_2$ ($a_+$) which has once been controversial. {\it
Quantitatively} everything seems OK for $B\ra D,D^*$ (taking into account the
uncertainty on $V_{cb}$).
Turning now to heavy to light semileptonic decays ($B, D\ra K, K^*, \pi,
\rho$), NRQM predictions are not very different from those of the heavy to
heavy
case. As a particular consequence,  the scaling
relations of eq. (\ref{scale}) follow straightforwardly. But NRQM goes
further, it predicts also relations between vector and
axial form factors, similar to those predicted for the heavy to heavy
case. These relations are not confirmed by experiment:  $\Gamma (D\ra K^*)/
\Gamma (D\ra K)$ is predicted twice too large.

The huge progress made by Isgur and Wise and by a sery of others
authors\cite{wis}, \cite{bj} was to demonstrate three things: 1) the above
quark
model predictions for heavy to heavy transitions at $q^2_{max}$
are in fact exact in QCD in the infinite quark
mass limit, 2) in the same limit exact relations between all the heavy to heavy
form factors at
$q^2\ne q^2_{max}$ for $0^-, 1^-\ra 0^-, 1^-$ can also be derived from QCD 3)
the scaling law in eq. (\ref{scale}) can also be derived on general grounds
from QCD, provided $\vert \vec q\vert\ll M_P$.
 Most current models would not satisfy the second set of  relations
(see for example ref. \cite{bj}). It has been shown that a careful {\it
relativistic} treatment of spins for states in motion allows to construct quark
models satisfying automatically these relations\cite{gif}.
\subsection{Relativity in the calculation of absolute magnitudes of form
factors}
\label{sub:rel}
With the advent of heavy quark symmetry (HQS) machinery, the appeal of quark
models is partly lost, since the above simple relations can be derived in an
exact QCD framework. The quark model has still the interest of offering a very
intuitive realisation of HQS. However the main interest is displaced towards
those features which HQS do not predict. These are: a) the absolute magnitudes
of the form factors away from the no recoil point, b) the corrections to HQS,
subleading in $1/m_Q$ ($m_Q$ is the heavy quark mass),
 which in HQS depend on a host of new arbitrary functions.
hskip -1 cm\begin{table}
\centering
\begin{tabular}{|c|c|c|c|c|}
\hline
Ref.& $f^+_K(0)$ & $V(0)$ & $A_1(0)$&$ A_2(0)$\\ \hline Lat. \cite{abada}
&$0.65
\pm 0.18$&$0.95 \pm 0.34$&$0.63 \pm 0.14$& $0.45 \pm 0.33$ \\ \hline
Lat. \cite{victor}-\cite{lubicz2}&$0.63 \pm 0.08$&$0.86 \pm 0.10$&$0.53 \pm
0.03$& $0.19 \pm 0.21$ \\ \hline
Lat. \cite{bes}-\cite{bes2}&$0.90\pm0.08$&$1.43\pm 0.45$&$0.83\pm
0.14$&
$0.59 \pm 0.14$ \\
&$\pm0.21$&$\pm0.49$&$\pm0.28$&$\pm0.24$\\ \hline
QM \cite{wsb}& $0.76$&$1.23$&$0.88$&$1.15$ \\ \hline
QM \cite{wisg}& $0.76-0.82$&$1.1$&$0.8$&$0.8$ \\ \hline
SR \cite{aos}& $0.6\pm 0.10$&$-$&$-$&$-$ \\
\hline
SR \cite{bbd}& $0.6^{+0.15}_{-0.10}$&$1.1\pm 0.25$&$0.5\pm0.15$&$0.6\pm0.15$ \\
\hline
Exp. \cite{e691}&$0.70 \pm 0.08$&$0.9 \pm 0.3 $&$ 0.46 \pm 0.05
$&$0.0 \pm 0.2 $\\
&&$\pm0.1$&$\pm0.05$&$\pm0.1$\\ \hline \hline Ref.& $A_1(q^2_{max})$ &
$V(0)/A_1(0)$ & $A_2(0)/A_1(0)$&$f^0(q^2_{max})$\\ \hline Lat. \cite{abada}
&$0.62
\pm 0.09 $&$1.50\pm 0.28$&$0.7 \pm 0.4 $&$0.93 \pm 0.13$ \\ \hline
Lat. \cite{lubicz2} &$0.77 \pm 0.20$ &$1.6\pm 0.2 $& $0.4 \pm 0.4$& $-$\\
\hline
Lat. \cite{bes2}&$1.27 \pm 0.16\pm0.31$&$1.99 \pm 0.22 \pm 0.33$&$0.7 \pm
0.16\pm
0.17$&$-$\\
 \hline
QM \cite{aw}& $-$&$1.9$&$0.8$&$-$ \\ \hline
QM \cite{ks}& $-$&$1.0$&$1.0$&$-$ \\ \hline
QM \cite{wsb}& $-$&$1.1$&$1.3$&$1.15$ \\ \hline
QM \cite{wisg}& $-$&$1.4$&$1.0$&$-$ \\ \hline
SR \cite{bbd}& $-$&$2.2\pm0.2$&$1.2\pm0.2$&$-$ \\ \hline
Exp. \cite{e691}& $0.54 \pm 0.06$& $-$ & $-$& $-$ \\
&$\pm 0.06$&&& \\ \hline Exp.
\cite{e653}& $-$&$2.00 \pm 0.33 $&$ 0.82\pm 0.23$& $-$ \\
&&$\pm0.16$&$\pm0.11$& \\
\hline
\end{tabular}
\caption{\it{Semileptonic form factors for $D \rightarrow
K$ and $K^*$. ``Lat.'' refers to lattice QCD, ``QM'' to quark models, ``SR''
to QCD sum rules and ``Exp.'' to experiment.}} \label{tab:final}
\end{table}
We will not consider point b) in detail since a) is the main challenge. It will
be sufficient to say a word about corrections to HQS relations at $q^2_{max}$.
They are known\cite{luke} to be of order in $1/m_Q^2$. They seem to be
dominated by the Dirac spinors relativistic corrections to the axial current.
There are also corections to all the currents fom the overlap of wave functions
which differ from 1 if the heavy quark masses are finite. But they are small,
entering through the expression $(1/m_Q-1/m_Q')^2$ due to a sort of
Ademollo-Gatto theorem.
The relativistic correction reduces the axial current contribution, in the
right direction to lessen the above mentioned discrepancy of NRQM for $\Gamma
(D\ra K^*)/ \Gamma (D\ra K)$, while its being very small for $\Gamma
(B\ra D^*)/ \Gamma (B\ra D)$ explains the good  NRQM prediction\footnote{In the
case of $D\ra K, K^*$ the relativistic ``corrections'' are more than
corrections, since the final meson are light. We use here these corrections to
get a qualitative hint, assuming that the first order gives the right trend}.
Let us concentrate on point a). We start from the observation that the naive
NRQM fails {\it qualitatively} in two ways: i) the slope of form factors seems
much too small at {\it small} $\vert\vec q \vert$ ($q^2 \sim q^2_{max}$) ii)
the
form factors have on the contrary a much too steep falloff at {\it large}
$\vert\vec q \vert$.
These facts have been well known since very long and have already been
discussed in ref. \cite{elec}.
The smallness of the predicted slope at $q^2_{max}$ can be appreciated in the
heavy quark limit by the prediction\cite{cw} of the factor $\rho^2$ (minus the
slope of the Isgur Wise function at the origin). It is unambiguously predicted
by the NRQM to be $\rho^2=m_d^2 R^2/2$, where $m_d$ is the spectator quark
constituant mass
and $R$ is a radius\footnote{The ground state wave function radius
for the harmonic oscillator potential.}, rather safely related to the
spectrum, which should not be very different from $R^2=6\,\hbox{GeV}^{-2}$
 in the heavy
quark limit. This gives something like $\rho^2=0.3$, in contrast to the roughly
measured $\rho^2 > 1.0$ from $B\ra  D^* l \nu$.
The too steep falloff at large $\vert\vec q \vert$, particularly dramatic in
$D\ra \pi$ and even more in $B\ra \pi, \rho$ where large momentum transfers
are kinematically allowed, depends  on the wave function and is particularly
striking for gaussian wave functions.
Of course this failure of the naive quark model is very serious: it is
qualitative and it concerns what seems to be the specific domain left to quark
models. How do they try to escape this disappointing situation ?
Apart from the sery of purely adhoc recipes used in ref.
\cite{wisg}\footnote{Notwithstanding these recipes the model of
ref.\cite{wisg} leads to an extremely small result for $B\ra \pi, \rho$,
see table \ref{tab:extrab},
because the falloff still remains gaussian.} we observe two main theoretical
trends:
\begin{enumerate}
\item[1)] {\it Nearest pole dominance}.
 Many authors simply renounce to
predict the form factors from the quark model except at one point ($q^2=0$ or
$q^2=q^2_{max}$) and prefer to use nearest pole dominance.
This is done for example in refs. \cite{aw} and \cite{wsb}.
We have already discussed in the subsection \ref{sub:nk}
the validity of the nearest pole dominance approximation. Let us
simply observe that it is not obvious to explain why one can combine these two
different approaches and assume that the quark model is still valid at some
$q^2$, while the nearest $t$-channel pole is dominating at that same point.
\item[2)]{\it Relativistic center-of-mass motion effects.} This is a very old
idea, yet there is not, up to now, a systematic relativistic treatment of
semileptonics decays. Rather several interesting ideas have been put forward,
which have still to be connected together.
\end{enumerate}
At large $q^2$, the Lorentz contraction effect has since long been known to
smoothen
the gaussian falloff\cite{lp}. This effect is included in the quark model of
ref. \cite{wsb} and explains why it does not predict too small values at
$q^2=0$
even for $B\ra \pi$ ($f^+_\pi(0)\sim 0.3$). However, in our opinion\cite{gif},
the relativistic boost of spins counterbalances this by large depression
factors and still leads to very small values\footnote{The authors of
 ref. \cite{wsb} omitted to consider this effect.}.
At small $q^2$, the latter relativistic effects of spin \cite{gif} seem
happily to enlarge $\rho^2$ with respect to its ``static'' value $m_d^2<\vec
r^2>/3$, where $m_d$ is the spectator quark constituant mass and the average
is taken over the
rest frames wave functions. Recently an important progress has been made by
Close and Wambach\cite{cw} who found a larger additional contribution to
$\rho^2$ due to the Lorentz transformation of the {\it spatial} wave function.
This effect seems specific to the situation where $m_d \ll m_Q$. It would not
be present with the simple Lorentz contraction prescription. Quantitative
predictions are however hampered, according to us, by the fact that $<\vec
r^2>$ cannot be identified with the non relativistic radius: it is submitted
to relativistic binding corrections of the same order as the ones under
discussion.
In conclusion, the introduction of relativity in the quark models seems
promising in many respects, but it would stand on a more solid ground for $b\ra
c$ decays than for $c\ra s$, where nearest pole dominance is not to be
excluded (see sections \ref{sec:qcd} and \ref{sec:lat}).
 Moreover, $D\ra
\pi$ and $B\ra \pi$ seem to escape the possibilities of quark models.
The numerical
predictions from quark models are reported, in comparison with others
results, in tables \ref{tab:final}, \ref{tab:largeurs} and \ref{tab:extrab}.
\section {QCD SUM RULES}
\label{sec:qcd}
We will not give too many details on QCD sum rules methods since they
have been discussed by Guido Martinelli \cite{gm} in his talk on leptonic decay
constants. For further details we refer to ref. \cite{cnp}. The theoretical
basis of QCD sum rules is totally rigourous: it incorporates analiticity,
asymptotic freedom, and non pertrubative effects are implemented through
vacuum expectation values of some operators: $<\bar q q>$,
$<G_{\mu\nu}G^{\mu\nu}>$ and  $<\bar q \sigma_{\mu\nu}G^{\mu\nu}q>$. However,
the practical use of QCD sum rules encounters important difficulties: 1) The
matching between the perturbative and non perturbative domains is by no means
trivial, and in practice it depends on a parameter usually labelled $s_0$ the
value of which is to some extent arbitrary;  2) the estimate of the vacuum
expectation values of the condensates are estimated from other applications of
QCD sum rules, but with large uncertainties; 3) there is a strong dependence on
the heavy quark mass.
Compared to quark models, QCD sum rules do not have any difficulty with Lorentz
covariance: the treatment is covariant from the start. Even more important,
this technique is by no means restricted to one value of $q^2$. It can be
applied to any $q^2$ except the vicinity of $q^2_{max}$ and is therefore the
{\it tool to study the $q^2$ dependence of the form factors}. To our
knowledge, this study has only been performed in refs. \cite{bbd} and
\cite{ball}. Their conclusion is, for $D$ as well as $B$ meson decays,
that the vector form factors dependence on $q^2$ is compatible with
the nearest vector meson pole dominance, but that {\it the axial currents do
not
show anywhere any effect of the nearest axial meson pole}. The latter
conclusion
is quite a surprise, in contradiction with lattice QCD as we shall see.
More work is needed here to understand this point thoroughly. The numerical
predictions from QCD sum rules are  reported in tables \ref{tab:final},
 \ref{tab:largeurs} and \ref{tab:extrab}.
\begin{figure}[t]   
    \begin{center} \setlength{\unitlength}{1truecm} \begin{picture}(6.0,6.0)
\put(-6.0,-9.0){\special{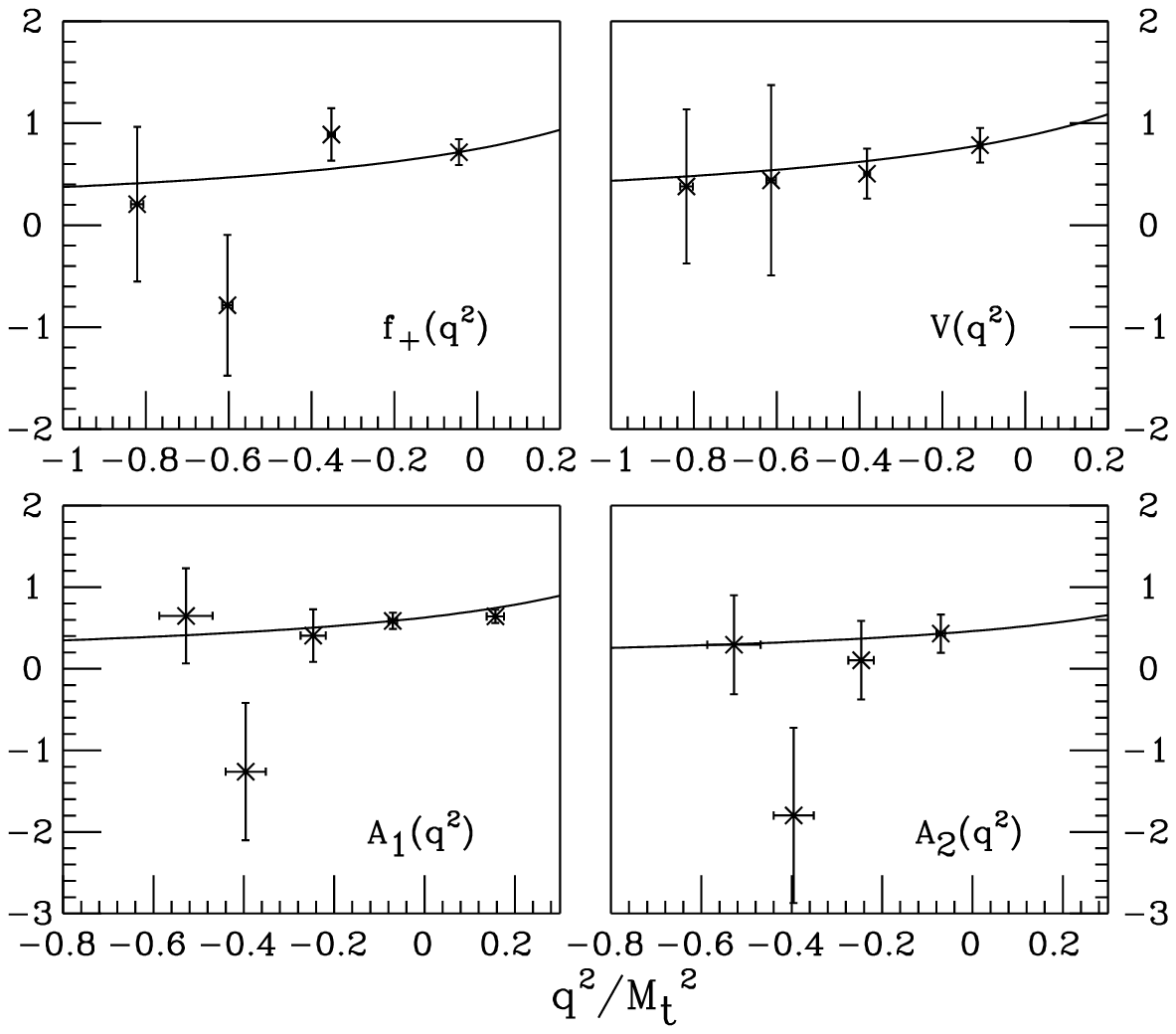}}
       \end{picture} \end{center} 	\vskip 2.6cm \caption[]{\it{We show
an example of the $q^2$ behaviour of the form factors. We have chosen quark
masses close to those involved in the $D\ra K, K^*$ decays.}}
\protect\label{md} \end{figure}
\section{EFFECTIVE LAGRANGIAN}
\label{sec:el}
The authors of ref. \cite{el} use an effective Lagrangian that incorporates
both heavy quark symmetry and chiral symmetry. Besides light ($q \bar q$) and
heavy-light ($Q \bar q$) pseudoscalar mesons, they incorporate light vector
mesons and heavy-light scalar mesons in their Lagrangian.  The parameters of
the model are tuned to $D\ra \pi$ and $D\ra K^*$, from which they predict $D\ra
\pi, \rho$ and $B\ra \pi, \rho, ...$. This is equivallent to extrapolating from
$D$ to $B$ through the scaling laws of ref. \cite{sliw},
neglecting the $O(1/m_Q)$ corrections, i.e. applying eq (\ref{scale}) with
$\delta_+=\delta_V=\delta_1=\delta_2=0$. The latter corrections may
nevertheless
not be negligeable as indicated by lattice calculations. The resulting form
factors
 are reported in table \ref{tab:extrab}.
\section{LATTICE QCD.}
\label{sec:lat}
Here again we will not repeat the general description of lattice QCD method by
Guido Martinelli.
Lattice QCD is based on first principles but suffers from several practical
limitations. Among these limitations, the fact that the cut-off (the inverse
lattice spacing) is rather low: $a^{-1}$ ranges from 2 to 4 GeV. The
consequence is that the masses of the ``heavy'' quark has to verify $m_Q\ll
a^{-1}$. This allows to study the charm quark, but for beauty an extarpolation
is needed, sometimes combined with direct data for infinitely  massive quarks.
Speaking of semileptonic decays, assuming that we work in the rest frame of the
initial meson, one has to vary $q^2$, i.e. $-\vec q$, the momentum of the final
meson. But the finite volume of the lattice allows only a discrete set of
values for the momentum: $\vec q=(n_x,n_y,n_z)2\pi/L$ where $L$ is the
spatial length of the lattice and $n_x$, $n_y$, $n_z$ are integers. In
practical calculations, $L$  ranges from 1 to 2 fermis. A further constraint is
that $\vert \vec q\vert \ll a^{-1}$. This means that only a few points are
computable, and the errors increase fast with $\vert \vec q\vert$.
Several calculations have been performed: refs. \cite{victor}-\cite{abada}.
Taking as an example the most recent one, ref. \cite{abada}, we notice that
the discreteness of momenta is not a big problem in the case of $D\ra K, K^*$,
because the parameters are such that the besides the zero momentum point
($q^2=q^2_{max}$), the point at  $\vec q=(1,0,0)2\pi/L$ happens to be located
close to $q^2=0$. Thus lattice calculations have direct access to both ends of
the physical region.
In figure \ref{md} taken from ref. \cite{abada} we compare the lattice points
to the nearest pole dominance, with the position of the pole computed from the
lattice. As can be seen, within large errors, lattice QCD is compatible with
nearest pole dominance in the $D\ra K, K^*$ decay. This seems to disagree, as
far as axial currents are concerned, with the results from QCD sum rules in
ref.
\cite{bbd}.
\begin{table}
\centering
\begin{tabular}{|c|c|c|c|c|}
\hline
$\vec p$ &$\gamma_+ \, \rm{GeV}^{-1/2}$ & $\gamma_V \, \rm{GeV}^{-1/2}$ &
$\gamma_1 \,\rm{GeV}^{+1/2}$& $\gamma_2 \, \rm{GeV}^{-1/2}$\\ \hline
$(0,0,0)$ &$-$&$-$&$0.96 \pm 0.16$& $-$ \\ \hline $(1,0,0)$ &$0.39 \pm
0.25$&$0.29 \pm 0.12$&$1.05 \pm 0.25$& $0.44 \pm 0.25$ \\ \hline \hline $\vec
p$& $\delta_+$ GeV & $\delta_V$ GeV & $\delta_1$ GeV &$\delta_2$ GeV\\ \hline
$(0,0,0)$ &$-$&$-$&$-0.33 \pm 0.09$& $-$ \\ \hline $(1,0,0)$ &$0.0 \pm
1.1$&$1.9 \pm 1.3$&$-0.46 \pm 0.22$& $-0.6 \pm 0.8$ \\ \hline
\end{tabular}
\caption{\it{The coefficients of the $1/m_Q$ expansion of the form
factors defined in eqs.(4).}}
\label{tab:scale}
\end{table}
\subsection{Extrapolation of lattice results to the B meson.}
\begin{figure}[t]   
    \begin{center} \setlength{\unitlength}{1truecm} \begin{picture}(6.0,6.0)
\put(-6.0,-9.0){\special{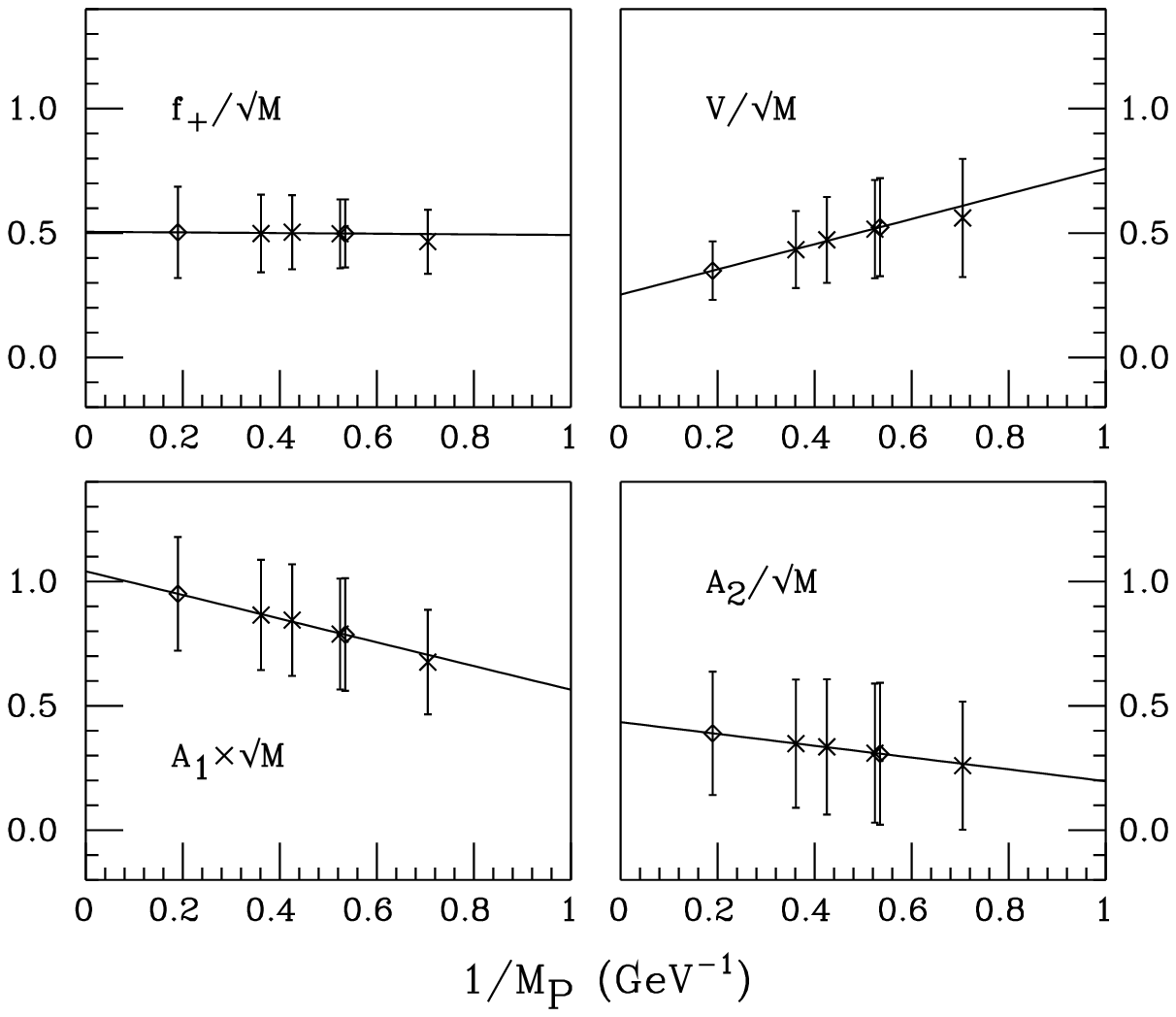}}
       \end{picture} \end{center}
\vskip 2.6cm
    \caption[]{\it{Form factors extrapolated to the chiral limit for the
light quark, as a function of the inverse pseudoscalar mass ($1/M_P$) for the
momentum assignment $\vec q=(2\pi/L,0,0)$.  The crosses are the lattice points,
the diamonds are the extrapolation to the D and B meson. Notice that the D is
very close to a lattice point.  The points corresponding to the lightest
heavy quark mass (furthest to the right) has not been used in the fits. }}
\protect\label{scaling}
\end{figure}

The extrapolation to $B$ meson needs two steps. i) For fixed $\vec q$, it
is possible from lattice, varying the heavy quark mass, to fit the parameters
of eq. (\ref{scale}). As an example we quote in table  \ref{tab:scale} the
results of ref. \cite{abada}, also illustrated in figure \ref{scaling}. Some
coefficients $\delta$ in table \ref{tab:scale}
 are large, indicating possible large corrections to scaling.
however, in view of the large errors, due to the limited
statistics in ref. \cite{abada}, it is impossible to draw any firm conclusion
except that {\it such an analysis is feasible with better statistics}.
 Soon, the statistics will improve,
thanks to dedicated computers. This step will lead to predictions for the $B$
meson form factors in the vicinity of $q^2_{max}$. ii)  The extrapolation of
the $B$ meson form factors down to $q^2=0$ raises yet unsolved theoretical
problems as we have argued in section \ref{sub:nk}.
Indeed, even if we take for granted from lattice results that the nearest pole
dominance is not a bad approximation in $D\ra K, K^*$, it does not allow to
assume its validity much further away from the pole, as is the case in $B\ra
\pi$ decay away from $q^2_{max}$. Remember that the range in $q^2$ is very
large and one cannot be satisfied with a rough extrapolation.
On the other hand, a {\it direct} access to
$B$ meson form factors near  $q^2=0$ needs tiny lattice spacing ($m_b\ll
a^{-1}$), i.e. a formidable increase in computer capacities.
\section{NUMERICAL COMPARISON OF THE RESULTS FROM DIFFERENT APPROACHES}
\subsection{$D$ meson decay.}
The results are summarised in tables \ref{tab:final} and \ref{tab:largeurs}.
All
the predictions concerning $f^+(0)$ are compatible with experiment. For
conserved
currents ($m_c=m_s$) $f^+(0)=1$. Quark model and VMD  help to understand why
for
$D\ra K$ the result is smaller than 1: For $q^2=q^2_{max}$ the mass difference
replaces the norm of the wave function, equal to 1, by an overlap of two
different wave functions, smaller than 1; the nearest pole dominance implies a
further decrease from $q^2_{max}$ down to $q^2=0$, further away from the pole.
In quark model $A_1$ is very similar to $f^+$ (the quark spin operator replaces
the identity operator). Lattice and QCD sum rules also predict $A_1$ not too
different from $f^+$, although the $A_1/f^+$ ratio is slightly
smaller in the latter models than in quark models. $V/A_1$ is rather large, 1
to 2 in all cases. $A_2/A_1$ is of order 1 in quark models and QCD sum rules,
smaller in lattice. For $A_2$ the situation is
unclear. E691 and ref.\cite{lubicz2} suggest a smaller value of $A_2$ than
E653, ref.\cite{bes} and \cite{abada}, even though the errors on this
quantity are so large that all the results are compatible.
Altogether, in view of the  errors still present in theory and experiment,
it is difficult to see more than an overall compatibility.  No clearcut
discrepancy is visible. It is worth here to stress that a better accuracy is
needed on both theoretical and experimental sides. For the latter, {\it a TCF
is really necessary}.
\begin{table}
\centering
\begin{tabular}{|c|c|c|c|}
\hline
Ref.& $\Gamma(D\rightarrow K)/ 10^{10}s^{-1}$ & $\Gamma(D\rightarrow K^*)/
10^{10}s^{-1}$ & $\Gamma(D\rightarrow \pi)/ 10^{10}s^{-1}$\\ \hline Lat.
\cite{abada} &$6.8\pm 3.4$&$6.0\pm 2.2 $&$0.56\pm 0.36 $\\ \hline
Lat. \cite{lubicz2}& $5.8\pm 0.15$&$5.0\pm0.9$&$0.5\pm0.2$ \\ \hline
exp. \cite{stone,ball} &$7.0 \pm 0.8$&$4.0\pm 0.7$&$0.9^{+0.5}_{-0.3}$ \\
\hline \hline Ref.&$ \Gamma(D\rightarrow \rho)/ 10^{10}s^{-1}$&
$\Gamma(D\rightarrow K^*)/\Gamma(D\rightarrow K)$ & $\Gamma_L/\Gamma_T$ \\
\hline Lat. \cite{abada} &$0.50\pm 0.23$& $0.92 \pm 0.55$&$1.27\pm 0.29$ \\
\hline Lat. \cite{lubicz2}& $0.4\pm 0.09$&$0.86\pm0.22$&$1.51\pm0.27$ \\ \hline
QM \cite {wsb}& $-$ & $1.14$&$0.89$ \\ \hline
QM \cite {wisg}& $-$ & $1.45$&$1.11$ \\ \hline
SR \cite {bbd}& $-$ & $0.5\pm0.15$&$0.86\pm0.06$ \\ \hline
Exp. \cite {pbu}&$-$& $0.63\pm0.09$&$-$\\ \hline
exp. \cite{stone} &$-$&$0.57\pm 0.08$&$1.15 \pm 0.17$ \\ \hline
\end{tabular}
\caption{\it{Semileptonic partial widths for $D \rightarrow K$, $K^*$,
$\pi$ and $\rho$, using $V_{cs}=0.975$ and $V_{cd}=0.222$.  We also report
the ratio of the longitudinal to transverse polarisation partial widths for
$D \rightarrow K^*$. The conventions in the first column are the same  as in
table 1.}} \label{tab:largeurs}
\end{table}
The small discrepancies add up and are enhanced in table \ref{tab:largeurs}
where
the partial widths are reported. The ratio
$\Gamma(D\rightarrow K^*)/\Gamma(D\rightarrow K)$ is predicted too large by
quark models, as we have already stated in section \ref{sub:iw}. QCD sum rules
predict a small number, while lattice are in between, with a large error.
\begin{table}
\centering
\begin{tabular}{|c|c|c|c|c|}
\hline
Ref.& $f^+(0)$ & $V(0)$ & $A_1(0)$&$ A_2(0)$\\ \hline Lat. \cite{abada} &$0.30
\pm 0.17$&$0.39 \pm 0.16$&$0.23 \pm 0.06$& $0.43 \pm 0.28$ \\ \hline
EL \cite{el} &$0.53 $&$0.62$&$0.21$& $0.20$ \\ \hline
QM \cite{wsb}& $0.33$&$0.33$&$0.28$&$0.28$ \\ \hline
QM  \cite{wisg} &$0.09$&$0.27$&$0.05 $& $0.02$ \\ \hline
SR \cite{ball} &$0.26 \pm 0.02$&$0.6\pm 0.2$&$0.5 \pm 0.1$&
$0.4 \pm 0.2$ \\ \hline
SR \cite{nar} &$0.23 \pm 0.02$&$0.47\pm 0.14$&$0.35 \pm 0.16$&
$0.42 \pm 0.12$ \\ \hline
\hline Ref.& $-$ & $V(0)/A_1(0)$ &
$A_2(0)/A_1(0)$&$-$\\ \hline
Lat. \cite{abada} &$-$& $1.5 \pm 0.6$&$1.9\pm 1.1$&
$-$ \\ \hline
\cite{wsb}& $-$ & $1.0$&$1.0 $ &$-$
 \\ \hline
\end{tabular}
\caption{\it{Semileptonic form factors for $B \rightarrow
\pi$ and $\rho$. The notations are the same as in table 1. ``EL'' refers to
effective Lagrangians.}} \label{tab:extrab}
\end{table}
\subsection{$B$ meson decay.}
Some results are reported in table \ref{tab:extrab}. One first remark is that
for most of the form factors the predictions of ref.\cite{wisg}
are much lower than all the others, cf. table \ref{tab:extrab}.  This results
in a much larger estimate of $\vert V_{ub}\vert$, for a given experimental
branching ratio. The reason for that has been discussed in section
\ref{sub:rel}: in ref.\cite{wisg} the form factor is
computed at $q^2_{max}$ and then a ``tempered'' exponential dependence on
$q^2$ is assumed. This $q^2$ yields a dramatic suppression at small $q^2$ for
B meson decays, where the range in $q^2$ is very large.
{}From eq. (\ref{scale}) the ratio $A_2/A_1$  scales like $M_P$ at fixed $\vec
q$.
The $q^2$ dependence being assumed to be rather smooth, we expect
$A_2(0)/A_1(0)$ to increase with $M_P$, although $\vert \vec q\vert\sim M_P/2$.
Table  \ref{tab:extrab}, compared with table \ref{tab:final} shows that indeed
lattice QCD predicts $A_2(0)/A_1(0)$ larger for $B$ decay than for $D$ decay.
Neither QCD sum rules, ref. \cite{ball}, nor the quark model of ref. \cite{wsb}
show this behaviour. Altogether table \ref{tab:extrab} shows a wide spreading
of
the predictions for the B semileptonic form factors.
This becomes even worst when one considers the partial widths. A summary is
given in table IV in ref. \cite{ball}. The outcome is that in units of $\vert
V_{ub}\vert^2 10^{13} s^{-1}$ the predictions for $\Gamma(B\ra \pi)$ range from
0.3 to 1.45 for QCD sum rules, from 0.21 to 0.74 for quark models,
to which we add the effective Lagrangian  of ref.\cite{el} prediction\footnote{
Concerning effective Lagrangian, we quote directly the numbers presented
by N. Di Bartolomeo in our working group. They differ from what is reported in
ref. \cite{ball}.}
of 1.9 and the
lattice prediction of ref.\cite{abada}: $1.2\pm 0.8$. For $\Gamma(B\ra \rho)$,
in the
same units, 0.77 to 3.3 from QCD sum rules,  1.63 to 2.6 from quark models,
2.12 from effective Lagrangian, and $1.3 \pm 1.2$ from lattice. The conclusion
is simple: {\it today, nothing reliable is known about the magnitude of $B$
meson
exclusive semileptonic decays}.
\section{USE OF THE FINAL HADRONS}
Although no work has yet been done to our knowledge on these issues, it should
be kept in mind that charm semileptonic decay $D\ra K^{**} l \nu$ is a clean
factory for $K^{**}$, where by $K^{**}$ we mean any excited strange meson,
heavier than $K^*(892)$. There would be a rich opportunity to increase our
understanding of strange meson spectroscopy if a TCF was to be built.
Next, the semileptonic decay $D\ra K\pi l
\nu$, with the $K\pi$ invariant mass below 892 MeV, is a unique, model
independent, access to $K\pi$ phase shift, analogous to what has been
 done about $\pi\pi$ phase shift from the  $K$ semileptonic decay into
two pions. This prospect has been repeatedly stressed by Jan Stern.
\section{CONCLUSION}
An accurate knowledge of CKM angles is necessary to understand the Standard
Model mass matrix, the generation puzzle. Such a knowledge needs a $b$-factory.
But without a theoretical knowledge of $B$ matrix elements, to the wanted
accuracy, a $b$-factory would be useless. There is no better testing band of
the
models, methods and ideas used to estimate these matrix elements,
than charm decays, whose relevant KM angles are already known to a sufficient
accuracy. If nature had provided us with a heavier charm, closer to the $b$,
life would be easier, we would not need such a distant extrapolation. But we
have no choice, we have to manage with the existing charm and to perform a long
distance extrapolation. We have to learn the corrections to scaling laws when
the heavy mass varies. We have to learn how the form factors depend on $q^2$.
This is the challenge to theorists. Under this proviso, {\it a TCF and a
$b$-factory appear not as competitors but as complementary facilities},
necessary to
reach both a better understanding of non perturbative QCD and and a better
 accuracy on CKM angles.


\begin{thebibliography}{999}
\bibitem{sliw}  N. Isgur and M.B. Wise, \prd{D42} (1990) 2388.
\bibitem{bsuv}I.I. Bigi, M. Shifman, N.G. Uraltsev and A. Vainshtein,
UND-HEP-93-BIG01.
\bibitem{aw}T. Altomari and L. Wolfenstein, \prl{58} (1987) 1583.
\bibitem{gilopr} E. Golowich, F. Iddir, A. Le Yaouanc, L. Oliver,
 O. P\`ene and J.C. Raynal, \pl{B213} (1988) 521.
\bibitem{wis}
 N.Isgur and M.Wise, Phys.Lett.B232 (1990) 113; Phys. Lett. B237 (1990) 527;
 Phys. ReV. Lett. 66 (1991) 1130;
H.Georgi and F.Uchiyama,  Phys. Lett. B238 (1990) 395; H.Georgi and
 M.Wise, Phys. Lett. B243 (1990) 279; J.D. Bjorken, 25th Int. Conf. on High
Energy
 Physics, Singapore, Aug 2-8, 1990, p 329;  H.Georgi, \np{B361} (1991) 339.
\bibitem{bj} J.D. Bjorken SLAC Summer Inst. 1990 p 167.
\bibitem{gif} A. Le Yaouanc et al., Gif lectures 1991 (electroweak properties
of heavy quarks), tome 1, p89.
\bibitem{luke}M.E. Luke, \pl{B252} (1990) 447.
\bibitem{elec}A. Le Yaouanc et al. \np{B37} (1972) 552.
\bibitem{cw}F.E. Close and A. Wambach, RAL-93-022, may-june 93.
\bibitem{wisg}N. Isgur, D. Scora, B. Grinstein and M.B. Wise, \prd{D39}
(1989) 799; N. Isgur and D. Scora \prd{D40} (1989) 1491.
\bibitem{wsb} M. Bauer, B. Stech and M. Wirbel,
Z. Phys. \underline{C29} (1985) 637; \underline{C34} (1987) 103.
\bibitem{lp} A.L. Licht and A. Pagnamamenta, \prd{D2} (1970) 1150 and 1156.
\bibitem{ks} K\"orner and Schuler Z. Phys. \underline{C38} (1988) 511;
\underline{C41} (1989) 690.
\bibitem{gm}{G.Martinelli, These proceedings.}
\bibitem{cnp}P. Colangelo,G. Nardulli and N. Paver BARI-TH/93-132, bulletin
board hep-ph/9303220.
\bibitem{bbd}P. Ball, V.M. Braun, H.G. Dosch \prd{D44} (1991) 3567.
 \bibitem{ball}P. Ball, TUM-T31-39-93, Bulletin Bd.:
hep-ph@xxx.lanl.gov - 9305267.
\bibitem{aos}T.M. Aliev, A.A. Ovchinnikov and V.A. Slobodenyuk, Trieste
Preprint IC/89/382 (1989).
\bibitem{el} R. Casalbuoni, A. Deandrea, N. Di
Bartolomeo, R. Gatto, F. Feruglio, G. Nardulli \pl{B299} (1993) 139.
\bibitem{victor} M.Crisafulli et al., Phys.Lett. \underline{223B}
(1989) 90.
\bibitem{lubicz} V.Lubicz, G.Martinelli and C.T.Sachrajda,
Nucl.Phys. \underline{B356} (1991) 310.
\bibitem{lubicz2} V.Lubicz, G.Martinelli, M.McCarthy and
C.T.Sachrajda, Phys.Lett. \underline{274B} (1992) 415.
\bibitem{bes} C.Bernard, A.El-Khadra and A.Soni,
Phys.Rev.\underline{D43} (1992) 2140..
\bibitem{bes2} C. Bernard, A. El-Khadra and A. Soni, Phys.Rev.
\underline{D45} (1992) 869.
\bibitem{abada}As. Abada et al. LPTENS 93/14.
\bibitem{nar}S. Narison, \pl{B283} (1992) 384.
\bibitem{e691} J.C. Anjos et al. \prl{65} (1990) 2630.
\bibitem{e653} K. Kodama et al. \pl{B274} (1992) 246.
\bibitem{stone} See for example S.Stone, in
Heavy Flavour Physics, eds. A.J. Buras and H. Lindner, world scientific,
Singapore (1992).
\bibitem{pbu} CLEO II, as reported by P. Burchat,
 5th International Symposium on Heavy Flavour Physics, Montreal, Canada, 6-10
July 1993, to appear in the proceedings.
\end{thebibliography}
\end{document}